\documentclass[nobibnotes,showpacs,showkeys]{revtex4}
\usepackage{amsfonts}
\usepackage[dvips]{color}
\usepackage{newlfont}
\usepackage{graphicx}
\usepackage{amsmath}
\usepackage[latin1]{inputenc}
\usepackage{float}

\begin{document}

\author{Adélcio C. Oliveira}
\email{adelcio@uesc.br}

\affiliation{Departamento de Ciências Exatas e Tecnológicas,
Universidade Estadual de Santa Cruz, Ilhéus, 45662 000, Bahia,
Brazil}

\author{A. R. Bosco de Magalhães}
\affiliation{Departamento de F{í}sica e Matemática, Centro Federal de Educa{çã}o Tecnoló%
gica de Minas Gerais, Belo Horizonte, MG, 30510-000, Brazil }

\title{Decoherence and Irreversibility: the role of the reservoir effective
Hilbert space size}

\begin{abstract}
We show that decoherence is determined by the effective Hilbert
space size, and demonstrate that a few degrees of freedom system can
simulate a \textquotedblleft\ N \textquotedblright\ degrees of
freedom environment if they have the same effective Hilbert space.
The effective Hilbert space size of the environment also determines
its long time efficiency as generator of irreversible coherence
loss.
\end{abstract}

\maketitle
\bigskip

\section{Introduction}

There are many situations where the direct application of the Schroedinger
equation leads to unsatisfactory results, a well known example is
dissipation. Dissipation can be phenomenologically modeled by two main
approaches, the complex Hamiltonian \cite{Golobtsova,Mensky} and open system
\cite{Caldeira,Diosi,Gesin}. It is also possible to quantize a
non-Hamiltonian system \cite{Tarasov}, but this approach is similar to the
open system one. The main idea is based on the fact that once we are dealing
with a physical phenomenon there are many particles involved. Due to the
difficulty of dealing with a large number of particles, it is common to
treat the system of interest as an open system, i.e., coupled with a
reservoir, which we may call the environment. It is a common belief that the
environment should be modeled by a system with infinity degrees of freedom.

While dissipation is a classical phenomena, in a real quantum system we have
also another effect, decoherence, which is an independent phenomena \cite%
{Romero}. The main idea of decoherence can be summarized as Omnes's \cite%
{Omnes1992} definition: \textit{``Decoherence is a dynamic effect through
which the states of the environment associated with different collective
states become rapidly orthogonal. It comes from a loss of local phase
correlation between the corresponding wave functions of the environment,
which is due to the interaction between the collective system and the
environment, also responsible for dissipation. It depends essentially upon
the fact that the environment has a very large number of degrees of
freedom.''\ }See also \cite{Zurek2003, Schlosshauer2004}. This loss of
coherence has been experimentally observed \cite{Haroche2001}.

For many practical situations the environment has been chosen as a
collection of harmonic oscillators \cite{Caldeira,Omnes1992,Zurek2003,
Schlosshauer2004,Isar,Oliveira06} and its phenomenological success has been
also proved in some cases \cite{Gardiner1994,Parkins1999,Bosco2004}. One of
the problems of quantum open systems is the difficulty of modeling the
environment by first principles and therefore its basis is phenomenological.
There are also many technical difficulties to obtain the so called master
equations and there is not a unique receipt \cite{Weiss,Bosco20042,Uzma}.
From the theoretical point of view there is a strong debate on the necessity
of using an environment in order to recover classical dynamics from Quantum
Mechanics. This is a subject of the classical limit problem \cite%
{Zurec98,ball04,Kofler,Bonifacio}. It was demonstrated that diffusion can
produce only quantum coherence attenuation, being not able to eliminate it
\cite{Oliveira06}, thus we are forced to include a finite experimental
resolution to reach the classical limit \cite{Oliveira08}.

Decoherence of a simple system can be quantified in terms of entropy, or
linear entropy \cite{Oliveira06,Kim96}. If the linear entropy of a single
system is zero we can say that it has not lost quantum coherence, while the
system energy has no direct relation with it. Usually the term decoherence
is applied to an irreversible loss of coherence of a quantum system, but can
we always measure the reversibility? Is it possible, for any practical or
philosophic reason, to use a few degrees of freedom system as an
environment? The main purpose of this work is to investigate these
questions. As a model, we use the quartic oscillator coupled with a phase
reservoir. The use of this model has two practical reasons, it has
analytical solutions and its classical limit is characterized by a unique
time \cite{renato2007,renato2006}, while in general we have many time scales
\cite{Oliveira03}. Also this model has experimental interest: it is used to
study Bose-Einstein condensate \cite{Greiner} and Kerr like medium \cite%
{Agarwal}.

\section{Phase Reservoir Model}

Let us consider a general phase reservoir Hamiltonian
\begin{equation}
H=H_{1}+H_{2}+H_{12},
\end{equation}
where $H_{1}$\ refers to the system of interest, $H_{2}$ to the
reservoir and $H_{12}$ to their interaction. The phase reservoir is
characterized by the commutation relations
\begin{equation}
\left[ H,H_{1}\right] =\left[ H,H_{2}\right] =\left[ H,H_{12}\right] =0.
\label{commrelations}
\end{equation}
Taking $\left| \phi _{n}\right\rangle $ and $\left| \psi _{k}\right\rangle $
as eigenvectors of $H_{1}$ and $H_{2}$, respectively, relations (\ref%
{commrelations}) permit us to write
\begin{equation}
H\left| \phi _{n}\right\rangle \left| \psi _{k}\right\rangle =E_{n,k}\left|
\phi _{n}\right\rangle \left| \psi _{k}\right\rangle .
\end{equation}
If $H_{2}$ concerns $M$ independent subsystems, we may assume

\begin{equation}
H_{2}=\sum_{l=1}^{M}H_{2}^{l},\qquad H_{12}=\sum_{l=1}^{M}H_{12}^{l},
\end{equation}%
with $\left[ H_{2}^{l},H_{2}^{k}\right] =\left[ H_{12}^{l},H_{12}^{k}\right]
=0$. The eigenvectors and eigenvalues will be given by $H_{2}^{l}\left\vert
\psi _{r_{l}}^{l}\right\rangle =E_{r_{l}}^{l}\left\vert \psi
_{r_{l}}^{l}\right\rangle $ and $H_{12}^{l}\left\vert \phi _{n}\right\rangle
\left\vert \psi _{r_{l}}^{l}\right\rangle =E_{n,r_{l}}^{1,l}\left\vert \phi
_{n}\right\rangle \left\vert \psi _{r_{l}}^{l}\right\rangle $. In order to
quantify the entanglement between the system of interest and the
environment, we may use the linear entropy $\delta (t)$. If $\rho (t)$ is
density operator for the whole system and $\rho _{1}(t)$ its trace over the
environmental variables, the linear entropy is calculated as $\delta
(t)=1-Tr\left\{ \left[ \rho _{1}(t)\right] ^{2}\right\} $. For the generic
initial state
\begin{equation}
\rho (0)=\sum_{v,w}B_{v,w}\left\vert \phi _{v}\right\rangle \left\langle
\phi _{w}\right\vert \prod_{l=1}^{M}\left\{
\sum_{r_{l},s_{l}}A_{r_{l},s_{l}}^{l}\left\vert \psi
_{r_{l}}^{l}\right\rangle \left\langle \psi _{s_{l}}^{l}\right\vert \right\}
,
\end{equation}%
we find
\begin{equation}
\delta (t)=1-\prod_{l=1}^{N}\left\{
\begin{array}{c}
\sum_{r_{l},s_{l}}%
\sum_{v,w}A_{r_{l},r_{l}}^{l}A_{s_{l},s_{l}}^{l}B_{v,w}B_{w,v} \\
\times \exp \left[ -\dfrac{it}{\hbar }\left(
E_{v,r_{l}}^{1,l}-E_{w,r_{l}}^{1,l}-E_{v,s_{l}}^{1,l}+E_{w,s_{l}}^{1,l}%
\right) \right]
\end{array}%
\right\} .
\end{equation}%
Notice that $\delta (t)$ depends only on the interaction part of the whole
Hamiltonian.

In order to analyze a specific case, we now assume that $H_{12}=\sum_{l}%
\lambda _{l}(\hbar N_{1})^{x}(\hbar N_{2}^{l})^{y}$, where $N_{1}\left| \phi
_{v}\right\rangle =v\left| \phi _{v}\right\rangle $ and $N_{2}^{l}\left|
\psi _{r_{l}}^{l}\right\rangle =r_{l}\left| \psi _{r_{l}}^{l}\right\rangle $
( $v$ and $r_{l}$ are integer numbers), leading to

\begin{equation}
\delta (t)=1-\prod_{l=1}^{M}\left\{
\begin{array}{c}
\sum_{r_{l},s_{l}}%
\sum_{v,w}A_{r_{l},r_{l}}^{l}A_{s_{l},s_{l}}^{l}B_{v,w}B_{w,v} \\
\times \exp \left\{ -it\hbar ^{x+y-1}\lambda
_{l}[(v^{x}-w^{x})(r_{l}{}^{y}-s_{l}{}^{y})]\right\}
\end{array}%
\right\} .  \label{delta geral}
\end{equation}%
The linear entropy behavior is usually characterized by decoherence time
\cite{Kim96} and revival time. Performing the expansion $\delta (t)\approx
\delta (0)+\delta _{1}t+\delta _{2}t^{2}+O(t^{3})$, it is easy to see that $%
\delta _{1}=0$. Thus, decoherence time is defined as $t_{D}=1/\sqrt{\delta
_{2}}$ and, for the system of interest in initially pure state ($\delta (0)=0
$), the explicit expression for this time reads
\begin{equation}
t_{D}=\frac{1}{\hbar ^{x+y-1}\Delta _{1}\sqrt{2\sum_{l=1}^{M}\left( \lambda
_{l}\Delta _{2}^{l}\right) ^{2}}},  \label{td geral}
\end{equation}%
where $(\Delta _{1}{}{})^{2}$ is $\left( N_{1}\right) {}^{x}$ operator
variance and $(\Delta _{2}^{l})^{2}$ is $(N_{2}^{l\text{ }})^{y}$ operator
variance for $\rho (0)$. Decoherence time (\ref{td geral}) depends solely on
the coupling Hamiltonian and the initial state of system plus environment.

Supposing that the coupling Hamiltonian frequency has a least multiple
frequency $\Lambda $, it recovers its purity at the revival time
\begin{equation}
t_{R}=2\frac{s\pi }{\hbar ^{x+y-1}\Lambda }.  \label{TR}
\end{equation}
In this case, we may define another important characteristic time, the
revival life time $\tau _{R}$, as the time interval when the recovered
purity can be observed. By expanding $\delta (t)$ around $t_{R}$, we obtain $%
\tau _{R}=2t_{D}$, which may be written as

\begin{equation}
\tau _{R}=\frac{\sqrt{2}}{\hbar ^{x+y-1}\Lambda \Delta _{1}\Delta _{2}},
\label{tau R geral}
\end{equation}%
where $\Delta _{2}=$ $\sqrt{\sum_{l=1}^{M}\left( \Delta
_{2}^{l}/k_{l}\right) ^{2}}$and $\ \ k_{l}=\Lambda /\lambda _{l}$. The
expressions for $t_{D}$ and $\tau _{R}$ indicate that decoherence process
can occur even for environments with few degrees of freedom: the crucial
factor is the product $\Delta _{1}\Delta _{2}$ compared to $1/\left( \hbar
^{x+y-1}\Lambda \right) $. From equations (\ref{td geral}) and (\ref{tau R
geral}) we can say that, in what concerns the main behavior of linear
entropy of the system of interest, the reservoir composed of $M$ systems is
phenomenologically equivalent to one degree of freedom with an effective
coupling constant $\Lambda $ and state variance $\left( \Delta _{2}\right)
^{2}$.

We get some insight by investigating the non-linear oscillator Hamiltonian:
\begin{equation}
H=\hbar \omega a^{\dagger }a+\hbar ^{2}g(a^{\dagger
}a)^{2}+\sum_{i=1}^{M}\hbar \Omega b_{i}^{\dagger }b_{i}+\sum_{j=1}^{M}\hbar
^{2}\lambda a^{\dagger }ab_{j}^{\dagger }b_{j},  \label{Hamilt}
\end{equation}
where $a^{\dagger }$ ($a$) and $b_{i}^{\dagger }$ ($b_{i}$) are creation
(annihilation) bosonic operators. This Hamiltonian fulfils the conditions
discussed above with $x=y=1$, and its eigenstates are the same as the ones
of the harmonic oscillator. Considering that the states of all environmental
bosons are initially the same, and defining the recurrence time $t_{r}$ as $%
\rho _{1}(t_{r})=\rho _{1}(0)$, we get: if $\lambda =0$, then $t_{r}=n\pi
/\hbar g$; else, for $g/\lambda \in
\mathbb{Q}%
$ we have $t_{r}=\frac{2\pi m}{\hbar \lambda n}s$, where $\frac{n}{m}=2\frac{%
g}{\lambda }$ and $s,n,m$ $\in
\mathbb{N}%
$.

\begin{figure}[th]
\includegraphics[scale=0.5]{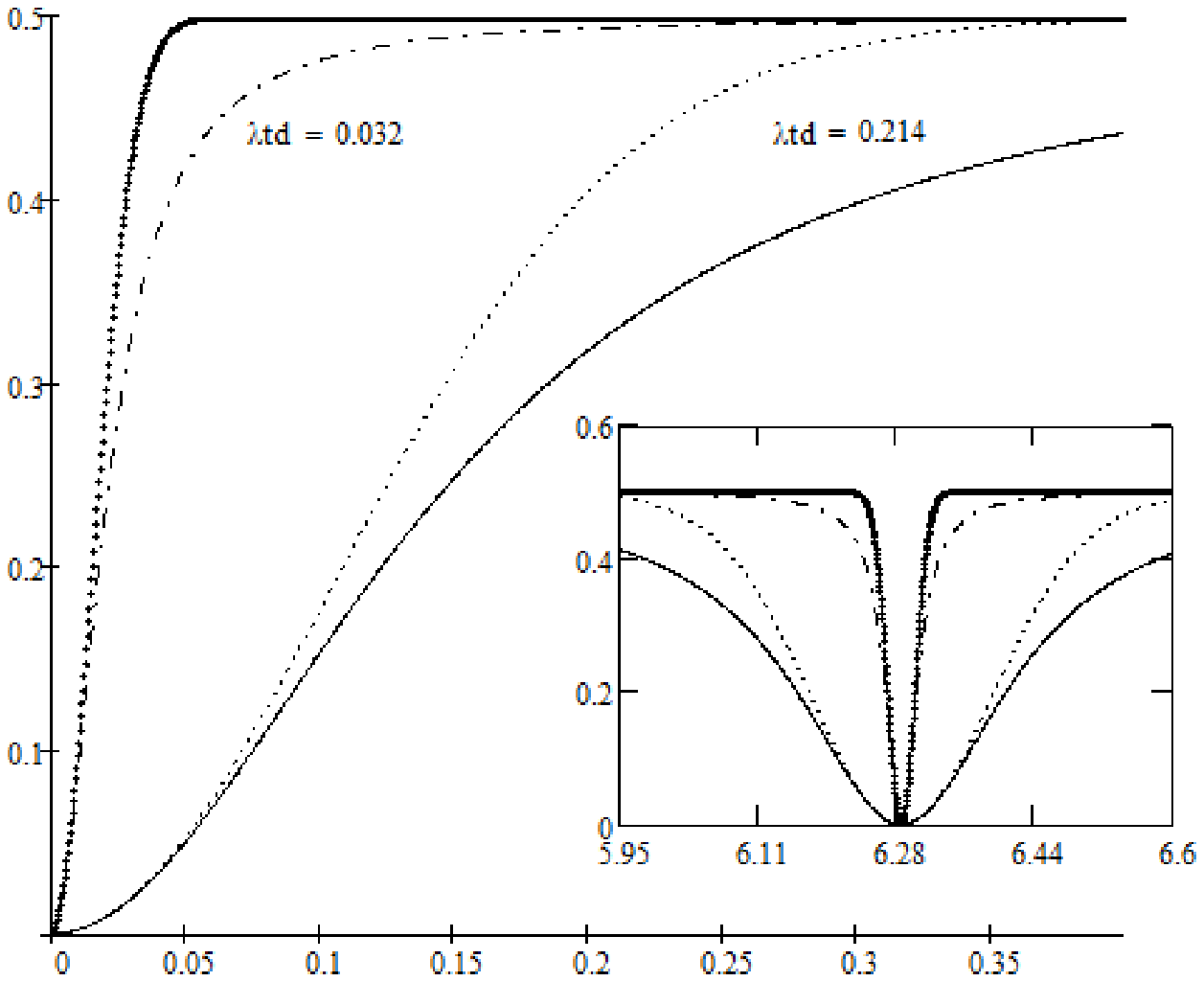} \vspace{-4.0cm}
\caption{It shows the linear entropy for thermal initial environmental
state. Thick dotted line: $M=201$, $\Delta _{2}=3.16$, $\protect\lambda %
t_{D}\approx 0.032 $. Dash-dotted line: $M=1$, $\Delta _{2}=44.83$, $\protect%
\lambda t_{D}\approx 0.032$. Full line: $M=1$, $\Delta _{2}=6.61$, $\protect%
\lambda t_{D}\approx 0.214$. Thin dotted line: $M=15$, $\Delta _{2}=1.71$, $%
\protect\lambda t_{D}\approx 0.214$. The insert show linear entropy around $%
t_{R}$. The horizontal axis corresponds to $\protect\lambda t$. }
\label{fig1}
\end{figure}

\begin{figure}[th]
\vspace{-2cm} \includegraphics[scale=0.5]{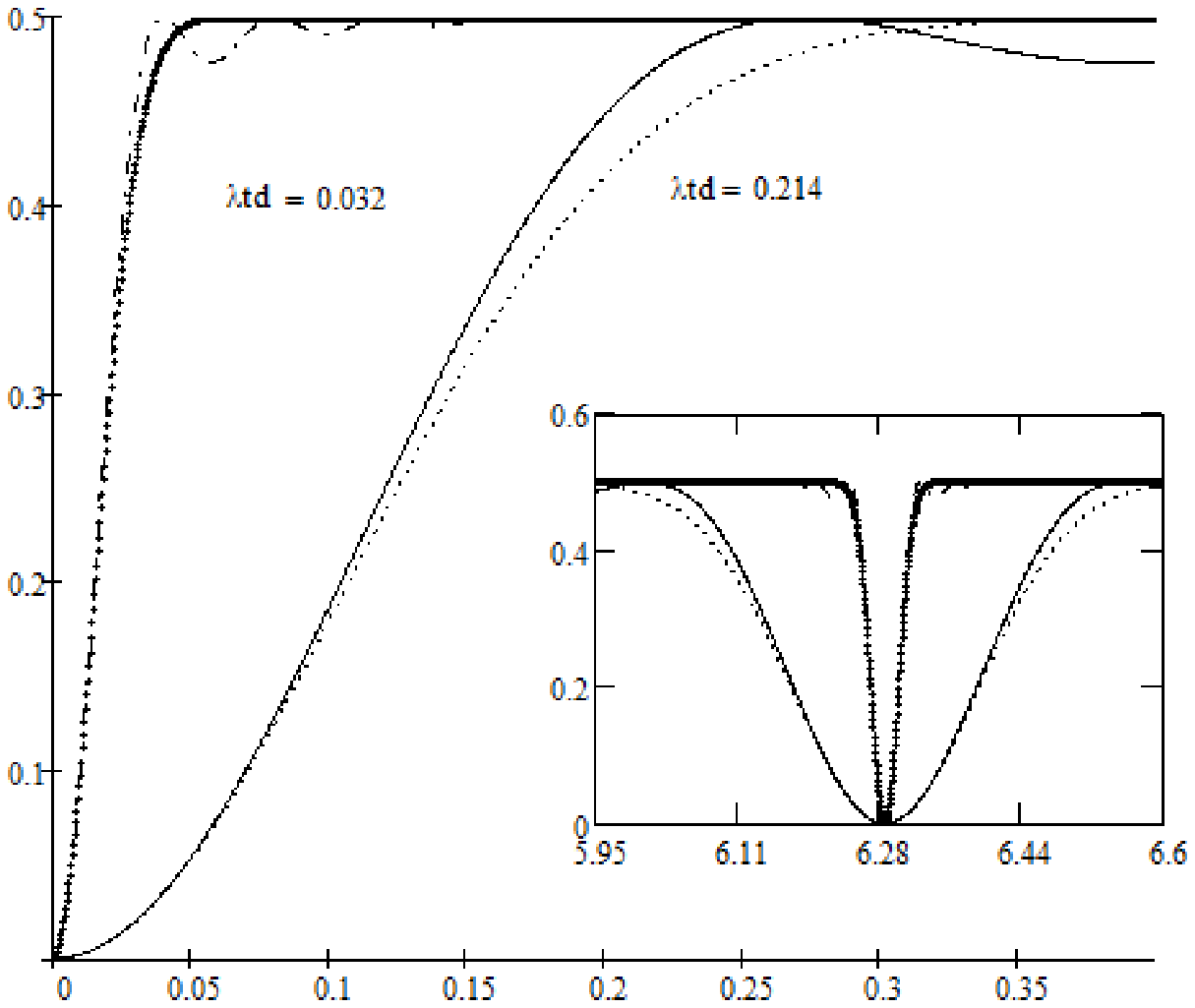} \vspace{-3cm}
\caption{Same as figure (\protect\ref{fig1}) for phase state reservoir with $%
m=0$.}
\label{fig2}
\end{figure}

From now on we assume the system of interest in initial state
\begin{equation}
\rho (0)=\frac{1}{2}\left\{ \left| 0\right\rangle \left\langle 0\right|
+\left| 1\right\rangle \left\langle 0\right| +\left| 0\right\rangle
\left\langle 1\right| +\left| 1\right\rangle \left\langle 1\right| \right\}
\otimes \prod_{k=0}^{M}r_{k}(0),  \label{R0}
\end{equation}
where $\left| 0\right\rangle $ and $\left| 1\right\rangle $ are Fock states
and $r_{k}(0)$ concerns the state of the $k$-th environmental oscillator.
Considering the reservoir in thermal equilibrium, i.e.,

\begin{equation}
r_{k}(0)=\left[ 1-\exp \left( -\frac{\hbar \Omega }{k_{B}T}\right) \right]
\sum_{n_{k}=0}^{\infty }\exp \left( -\frac{\hbar \Omega }{k_{B}T}%
n_{k}\right) \left\vert n_{k}\right\rangle \left\langle n_{k}\right\vert ,
\end{equation}%
where the $\left\vert n_{k}\right\rangle $ are Fock states, $k_{B}$ is
Boltzmann's constant and $T$ the absolute temperature, we obtain
\begin{equation}
\delta (t)=\frac{1}{2}\left\{ 1-\left[ \frac{\left[ 1-\exp \left( -\frac{%
\hbar \Omega }{k_{B}T}\right) \right] ^{2}}{1+\exp \left( -\frac{2\hbar
\Omega }{k_{B}T}\right) -2\exp \left( -\frac{\hbar \Omega }{k_{B}T}\right)
\cos (\lambda \hbar t)}\right] ^{M}\right\} ,
\end{equation}%
which explicitly shows that even for a large $M$ the system recovers purity
at $t=t_{R}$. In Fig. (\ref{fig1}), we show the linear entropy for this
thermal initial condition \footnote{%
For all figures we have chosen $\lambda =0.1$, $g=1$ and $\hbar =1$.}. The
insert shows the linear entropy around $t_{R}$: as we can observe, the
quantum behavior is attenuated as the total number of oscillators is
increased, as expected, and also at higher temperatures. It is clear from
this figure that a unique oscillator at high temperature and $M$ oscillators
with low energy are phenomenologically equivalent as an environment (since $%
\Delta _{2}$ increases with the temperature). In order to stress the fact
that (for fixed $\Delta _{2}$) the main features of the evolution of linear
entropy is independent of the specific reservoir initial state, we plotted
in Fig. (\ref{fig2}) the evolution of $\delta (t)$ for the reservoir in
Pegg-Barnett phase state \cite{Pegg} $\left\vert \varphi _{m}\right\rangle
(r)$, defined as

\begin{equation}
\left| \varphi _{m}\right\rangle (r)=\frac{1}{\sqrt{r+1}}\sum_{n=0}^{r}e^{in%
\varphi _{m}}\left| n\right\rangle ,  \label{phastate}
\end{equation}
\vspace{0cm}
\vspace{0cm} where $\varphi _{m}=\frac{2\pi m}{r+1},$ $m=0,1,...,r$.

The Pegg-Barnett phase state has a uniform distribution over the
lowest number states; also they form a truncated $r+1$ dimensional
Hilbert space, with $\Delta _{2}=\left[ r\left( r+2\right)
/12\right] ^{1/2}$. That decoherence properties of the evolutions of
Fig. (\ref{fig2})\ depend on the quantity $\Delta _{1}\Delta _{2}$.
For other initial states, we see, by the characteristics times
calculated, that decoherence properties depends also on this
quantity. Based on that, we define the \textit{effective Hilbert
space size} as the truncated Hilbert space size of the equivalent
phase state that generates the same purity loss of the reservoir
state considered:
\begin{equation}
Hs=\sqrt{1+12\left( \Delta _{2}\right) ^{2}}.  \label{HS}
\end{equation}
If $y$ $\neq 1$, we can use the same receipt: for each specific $N_{2}^{y}$
variance, there is an equivalent $r$ which defines $Hs$.
\vspace{0cm}

Now let us say a word about quantum uncertainty and reversibility. The above
result shows that the revival mean time $\tau _{R}$ goes to zero in the
limit $\Delta _{2}\longrightarrow \infty $. The important point that shall
be stressed is that this limit can be obtained even for an environment with $%
M=1 $, i.e., a one system reservoir in infinity temperature. In fact we have
not infinity temperature in laboratories. However, measuring this revival
for high temperatures (large $(\Delta _{2})^{2}$) demands high time
precision. The time uncertainty is related to the system energy uncertainty
by Heisenberg uncertainty principle as $\Delta t\Delta E\geq \hbar /2$.
Considering our initial state (\ref{R0}), and as we are interested in
revival dynamics, we get $\Delta t\geq \frac{1}{\lambda \hbar }$. For $%
\Delta t>>$ $\tau _{R}$ , or $\frac{\sqrt{2}}{\Delta _{1}\Delta _{2}}<<1$,
we would have a vanishing probability of observing the revival. The
Heisenberg uncertainty relation gives a fundamental limitation in
experimental resolution; in real experiments, time uncertainty is always
greater \cite{Quantummesurement}. This suggests a classical limit based on
coarse-grained measurement \cite{Bonifacio,Kofler}.

\begin{figure}[th]
\vspace{-2.5cm} \includegraphics[scale=0.5]{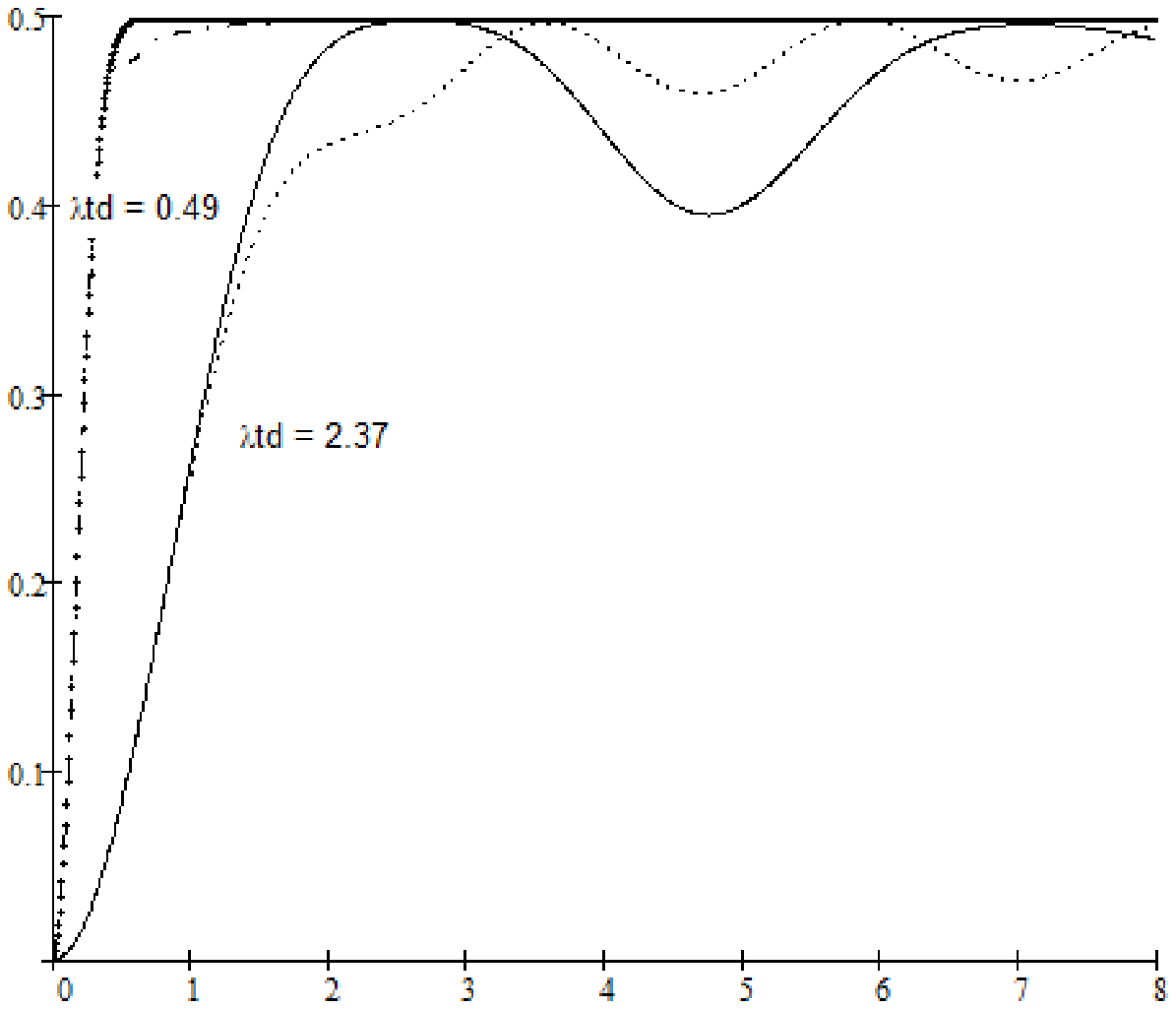} \vspace{-3.8cm}
\caption{It shows the linear entropy for phase state reservoir with $m=0$.
Thick dotted line: $r=10$, $M=20$, $\protect\lambda t_{D}\approx 0.49$.
Dash-dotted line: $r=289$,$\ M=1$, $\protect\lambda t_{D}\approx 0.49$. Full
line: $r=2$, $M=2$, $\protect\lambda t_{D}\approx 2.37$. Thin dotted line: $%
r=8$,$\ M=1,$ $\protect\lambda t_{D}\approx 2.37.$ In all graphics we used $%
y=1/2$. The horizontal axis corresponds to $\protect\lambda t$.}
\label{fig4}
\end{figure}

Another way for revival vanishing is to assume that $x$ or $y$ are not
integers. In figure (\ref{fig4}), we show the linear entropy evolution for
initial phase state and a new coupling term: $H_{12}=$ $\sum_{j=1}^{M}\hbar
^{2}\lambda a^{\dagger }a\sqrt{b_{j}^{\dagger }b_{j}}$. The revivals are
absent and partial purity recover is attenuated as we increase $\Delta _{2}$%
. Again we see that decoherence depends almost solely on $\lambda $ and $%
\Delta _{2}$.

\vspace{-0.1cm}

\section{Conclusion}

We showed that decoherence is determined by the effective Hilbert
space size (\ref{HS}) characterized by the reservoir state variance
and the coupling Hamiltonian. Thus the specific reservoir state is
not relevant for linear entropy dynamics. Also we showed that a
simple system can be an effective reservoir. Uncertain relations are
invoked to demonstrate that recoherence cannot be measured in a poor
resolution regime.

\textbf{Acknowledgments:} A.C.O. is acknowledge FAPESB for partial
financial support. A.R.B.M. acknowledge FAPEMIG (Process
APQ-2347-5.02/07) for partial financial support. We acknowledge M.
C. Nemes for fruitful discussions.

\end{document}